\documentclass{pasa}

\usepackage{graphicx}

\usepackage{tikz} 
\usepackage{amsmath}
\usepackage{hyperref}
\usepackage{ wasysym}
\usepackage{pgfplots}
\usepackage{wrapfig}
\usepackage{graphicx}

\usepackage{xcolor} 
\usepackage[normalem]{ulem} 

\usepackage{cuted}
\title[Parallactic delay for geodetic VLBI and non-orthogonality of the fundamental axes]{Parallactic delay for geodetic VLBI and non-orthogonality of the fundamental axes}

\author[O. Titov et al.]{O. Titov$^1$, and A. Osetrova$^{2}$
\affil{$^1$Geoscience Australia, Canberra, Australia}%
\affil{$^2$Saint-Petersburg State University, Saint-Petersburg, Russia}

}%

\jid{PASA}
\doi{10.1017/pas.\the\year.xxx}
\jyear{\the\year}

\usepackage{aas_macros}
\usepackage{hyperref} 
\hypersetup{colorlinks,citecolor=blue,linkcolor=blue,urlcolor=blue}

\pgfplotsset{compat=1.18}
\begin{document}
\begin{frontmatter}

\newcommand{\RomanNumeralCaps}[1]
    {\MakeUppercase{\romannumeral #1}}
\maketitle

\begin{abstract}

The Gaia optical astrometric mission has measured the precise positions of millions of objects in the sky, including extragalactic sources also observed by Very Long Baseline Interferometry (VLBI). In the recent Gaia EDR3 release, an effect of negative parallax with a magnitude of approximately $-17$~$\mu$as was reported, presumably due to technical reasons related to the relativistic delay model. A recent analysis of a 30-year set of geodetic VLBI data (1993–2023) revealed a similar negative parallax with an amplitude of $-15.8 \pm 0.5$ $ \mu$as. Since both astrometric techniques, optical and radio, provide consistent estimates of this negative parallax, 
it is necessary to investigate the potential origin of this effect. 

We developed the extended group relativistic delay model to incorporate the additional parallactic effect for radio sources at distances less than 1 Mpc, and found that the apparent annual signal might appear due the non-orthogonality of the fundamental axes, which are defined by the positions of the reference radio sources themselves. Unlike the conventional parallactic ellipse, the apparent annual effect in this case appears as a circular motion for all objects independently of their ecliptic latitude. The measured amplitude of this circular effect is within a range of 10--15 $\mu$as that is consistent with the ICRF3 stability of the fundamental axis. This annual circular effect could also arise if a Gödel-type cosmological metric were applied, suggesting that, in the future, this phenomenon could be used to indicate global cosmic rotation.

\end{abstract}
\end{frontmatter}

\section{INTRODUCTION }

 The Very Long Baseline Interferometry (VLBI) technique measures the time difference (time delay) between the arrivals of a radio signal from a distant source at two radio telescopes separated by a 
long baseline (\cite{Schuh12}). This time delay is calculated by correlating the signals recorded at both stations, using either hardware or software correlators. By observing many radio sources over extended periods, highly accurate position estimates for these sources can be achieved. 

 Two previous fundamental catalogs of reference radio sources, ICRF1 and ICRF2, were released in 1995  (\cite{Ma1998})  and 2009  (\cite{Fey_2015}),  respectively. The current International Celestial Reference Frame (ICRF3) is based on the positions of 4,536 radio sources observed between 1979 and 2018, of which 303 are "defining" sources that define the fundamental axes with an uncertainty of 30~$\mu$as (\cite{Charlot_2020}).  The formal errors of individual radio source positions vary widely due to the uneven number of observations, ranging from 6 $\mu$as to a few mas. 

 The relativistic model for the VLBI group delay, as published in the IERS Conventions 2010  (\cite{iers10}),  accounts for the relativistic effects of both special and general relativity with a precision of 1~ps. However, this model does not include the parallactic delay, as almost all radio sources observed by geodetic VLBI are located beyond 1 Mpc. A few Galactic radio stars are treated similarly to extragalactic objects  (\cite{Lunz_2023}). Recently, \cite{Soffel_2017} extended the conventional model to include closer objects, but this advanced model has not yet been adopted by the IERS Conventions. 

 In the recent Gaia EDR3 release, a negative annual parallax with an amplitude of approximately -17 $\mu$as was reported for extragalactic active galactic nuclei (AGNs) observed in optical wavelengths  (\cite{Lindegren+2021}).  The authors suggested that this effect might be due to the design of the Gaia telescope, which introduces calibration challenges related to the angle between its two mirrors, thus leading to the zero parallax point offset (e.g. \cite{Groenewegen_2021}). Accurate parallax measurements are crucial for various applications  (\cite{Butkevich_2017}),  making additional verification of this estimate essential. 

 Since nearly all radio sources observed by geodetic VLBI are at extragalactic distances, their parallax has long been considered to be  negligible and has therefore never been estimated. However, (\cite{Titov+2024}) recently identified a similar negative annual parallax of $-15.8 \pm 0.5$~ $\mu$as, based on 30 years of geodetic VLBI observations from 1993 to 2023. New CRF solution (aus2024a.crf\footnote{\label{fnote1}\url{https://cddis.nasa.gov/archive/vlbi/ivsproducts/crf/}}) enlarges the number of group delays to 15 millions, i.e. approximately on 2 millions more than used for ICRF3 catalogue (\cite{Charlot_2020}). A new catalogue comprises coordinates of 5428 radio sources with three or more observations, i.e. almost 1000 more than in ICRF3. The formal errors of the most observed radio sources is about 2-3 $\mu$as. Such progress encourages the investigation of a hidden systematic that could be missed in the previous solutions.
 
 It remains unclear whether the optical and radio instruments are detecting the same phenomenon or if this is merely a numerical coincidence between two independent effects.  

 To address this ambiguity, we develop an analytical VLBI delay model that includes parallax and explore possible origins for an annual signal that is independent of the distance to the observed objects. 
 
\section{PARALLACTIC DELAY}

For all extragalactic radio sources at a distance $R$ greater than 1 Mpc, the annual parallax is less than 1 $\mu$as, and the corresponding parallactic delay is negligible. Consequently, the equation for the parallactic delay has not been formulated until 2017 (\cite{Soffel_2017}). In this work, we develop the equation for the parallactic delay in a form that is convenient for implementation in coding and consider alternative of the annual signal appearance apart from the classical astrometric explanation. 

\begin{figure}[h!]
\begin{center}
\includegraphics[scale=0.6]{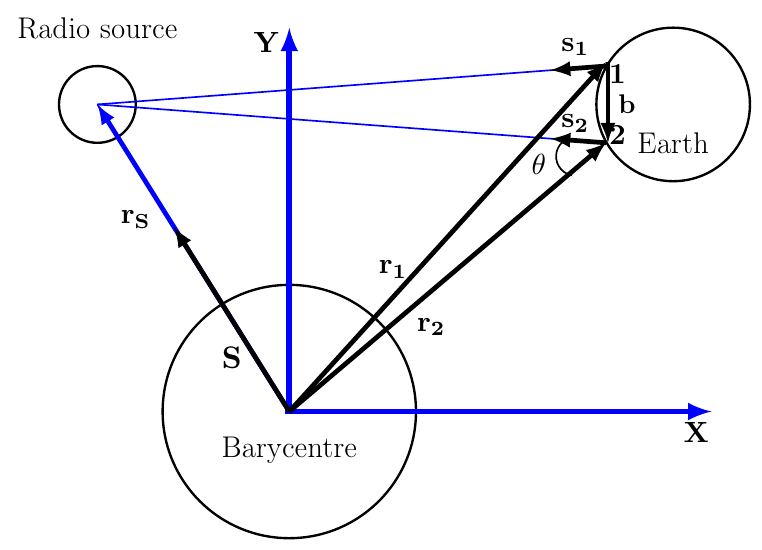} 
\vspace{0.1 cm}
\caption{Mutual positions of two radio telescopes on Earth and a radio source in the barycentric reference frame.}
\label{Fig1}
\end{center}
\end{figure}

The standard approach involves considering the difference between the barycentric coordinates of two radio telescopes, $\boldsymbol{r_1}(t_1)$ and $\boldsymbol{r_2}(t_2)$, to represent the difference in time propagation  $\tau$ as follows (Fig. \ref{Fig1}).
 
\begin{equation}\label{delay0}
\begin{aligned}
\tau = {t_2} -{t_1} =  {t_{2S}} + {t_{2E}} - ({t_{1S}} +{t_{1E}}) 
\end{aligned}
\end{equation}
Where   
\begin{equation}\label{delay}
\begin{aligned}
\tau = \frac{\boldsymbol{r_{S}} \boldsymbol{s_{2}} }{c} - \frac{\boldsymbol{r_{2}} \boldsymbol{S}}{c} - \Bigg(\frac{\boldsymbol{r_{S}} \boldsymbol{s_{1}} }{c} - \frac{\boldsymbol{r_{1}} \boldsymbol{S}}{c}\Bigg) =\\
= \frac{\boldsymbol{r_{S}} \boldsymbol{s_{2}} }{c} - 
\frac{\boldsymbol{r_{S}} \boldsymbol{s_{1}} }{c}
- \Bigg(\frac{\boldsymbol{r_{2}} \boldsymbol{S}}{c}  - \frac{\boldsymbol{r_{1}} \boldsymbol{S}}{c}\Bigg) = \\ 
= \tau_{S} + \tau_{E}
\end{aligned}
\end{equation}

The first term in equation (\ref{delay}) represents the time delay propagation from an extragalactic radio source to the Solar System barycenter, while the second term accounts for the local Solar System effects.
If we ignore the parallactic effect, then ${\tau_{S}} = 0$, and the total delay (\ref{delay0}) is reduced to the geometric delay $\tau_{E}$  in its simplest form. Denoting the vector baseline as $\boldsymbol{b} = \boldsymbol{r_{2}} - \boldsymbol{r_{1}}$:

\begin{equation}\label{baseline_delay}
\begin{aligned}
\tau = \tau_{E} = -\frac{(\boldsymbol{r_2} - \boldsymbol{r_1})\boldsymbol{S}}{c} = -\frac{(\boldsymbol{b}\boldsymbol{S})}{c}
\end{aligned}
\end{equation}

Equation (\ref{baseline_delay}) presents the geometric delay in its simplest form. In the full form it expands to the relativistic delay including all effects of special relativity and the retarded baseline correction due to the geocentric motion of the both radio telescopes. It is fully consistent to the official IERS Conventions 2010 model (\cite{iers10}), therefore, any additional terms developed in the paper, may be simply added to the conventional equation. Typically, only the expanded version of equation (\ref{baseline_delay}) is used for modeling geodetic VLBI observations of extragalactic objects. However, if we wish to account for the parallactic effect in the time delay, the first term ${\tau_{S}}$ in equation (\ref{delay}) must be elaborated. Throughout the manuscript, all additional terms will be developed in the same way as the conventional equations in previous papers (\cite{Titov_2015}, \cite{Titov_2020}).

In the general case, the unit direction vectors from each radio telescope to an extragalactic object differ (i.e. $\boldsymbol{s_1} \neq \boldsymbol{s_2}$), and the equation for $s_{i}$ (where $i = 1,2$)  is given by:

\begin{equation}\label{s_i}
\begin{aligned}
\boldsymbol{s_{i}} =\frac{\boldsymbol{r_{is}}}{|\boldsymbol{r_{is}}|} = \frac{\boldsymbol{r_S} - \boldsymbol{r_i}}{ 
\sqrt{\boldsymbol{r_S}^2 
 - 2(\boldsymbol{r_S}\boldsymbol{r_i}) + \boldsymbol{r_i}^2}}
\end{aligned}
\end{equation}

and, as for an extragalactic radio source $r_{S} \gg r_{i}$:

\begin{equation}\label{s_i2}
\begin{aligned}
\boldsymbol{s_{i}} = \frac{\boldsymbol{r_S} - \boldsymbol{r_i}}{|\boldsymbol{r_S}|
\sqrt{1-\frac{2(\boldsymbol{r_S}\boldsymbol{r_i})}{\boldsymbol{r_S}^2}+\frac{\boldsymbol{r_i}^2}{\boldsymbol{r_S}^2}}}
\end{aligned}
\end{equation}

Then, using the well-known expansion

\begin{equation}\label{math_decomposition}
\begin{aligned}
\frac{1}{\sqrt{1+x}} \approx 1 - \frac{x}{2} + \frac{3x^2}{8} - \frac{5x^3}{16} +...
\end{aligned}
\end{equation}

one could convert (\ref{s_i2}) as follows:


\begin{equation}\label{s_i3}
\begin{aligned}
\boldsymbol{s_{i}} & 
\approx \frac{\boldsymbol{r_S} - \boldsymbol{r_i}}{|\boldsymbol{r_S}|} 
\Bigg(1-\frac{1}{2}\Bigr(\frac{\boldsymbol{r_i}^2}{\boldsymbol{r_S}^2} - 
\frac{2({\boldsymbol{r_i}\boldsymbol{r_S}})}{\boldsymbol{r_S}^2}\Bigl) + \\
&+\frac{3}{8}
\Bigr(\frac{\boldsymbol{r_i}^2}{\boldsymbol{r_S}^2} - \frac{2({\boldsymbol{r_i}\boldsymbol{r_S}})}{\boldsymbol{r_S}^2}\Bigl)^2 \Bigg)
\end{aligned}
\end{equation}

By defining $\boldsymbol{S} = \frac{\boldsymbol{r_S}}{|\boldsymbol{r_S}|}$ and truncating terms of the order $O\Bigr(\frac{\boldsymbol{r_i^3}}{\boldsymbol{r_S^3}}\Bigl)$ and higher, one can obtain:

\begin{equation}\label{s_i4}
\begin{aligned}
\boldsymbol{s_{i}} \approx \boldsymbol{S}- & \frac{\boldsymbol{S}}{2}\frac{\boldsymbol{r_i}^2}{\boldsymbol{r_S}^2}  + 
\frac{(\boldsymbol{r_i}\boldsymbol{S})\boldsymbol{S}-\boldsymbol{r_i}}{|\boldsymbol{r_S}|} + 
\frac{3(\boldsymbol{r_i}\boldsymbol{S})^2\boldsymbol{S}}{2|\boldsymbol{r_S}|^2}  -
\frac{(\boldsymbol{r_i}\boldsymbol{S})\boldsymbol{r_i}}{|\boldsymbol{r_S}|^2} = \\
& =  \boldsymbol{S} +
\frac{(\boldsymbol{r_i}\boldsymbol{S})\boldsymbol{S}-\boldsymbol{r_i}}{|\boldsymbol{r_S}|} -  \\
& - \frac{1}{|\boldsymbol{r_S}|^2}\Bigg((\boldsymbol{r_i}\boldsymbol{S})\boldsymbol{r_i}
+\frac{\boldsymbol{S}\boldsymbol{r_i}^2}{2}-\frac{3(\boldsymbol{r_i}\boldsymbol{S})^2\boldsymbol{S}}{2}\Bigg) 
\end{aligned}
\end{equation}

Therefore, for $(\boldsymbol{r_S}\boldsymbol{s_i})$:

\begin{equation}\label{dot_product}
\begin{aligned}
&(\boldsymbol{r_S}\boldsymbol{s_i})  \approx 
(\boldsymbol{r_S}\boldsymbol{S}) +
\frac{(\boldsymbol{r_i}\boldsymbol{S})(\boldsymbol{r_S}\boldsymbol{S})-(\boldsymbol{r_S}\boldsymbol{r_i})}{|\boldsymbol{r_S}|} - \\
&- \frac{1}{|\boldsymbol{r_S}|^2}
\Bigg((\boldsymbol{r_i}\boldsymbol{S})(\boldsymbol{r_S}\boldsymbol{r_i})
+\frac{(\boldsymbol{r_S}\boldsymbol{S})\boldsymbol{r_i}^2}{2}-\frac{3(\boldsymbol{r_i}\boldsymbol{S})^2(\boldsymbol{r_S}\boldsymbol{S})}{2}\Bigg) 
\end{aligned}
\end{equation}

As $(\boldsymbol{r_S}\boldsymbol{S}) = |\boldsymbol{r_S}|$, the second term in (\ref{dot_product}) disappears, and

\begin{equation}\label{dot_product2}
\begin{aligned}
(\boldsymbol{r_S}\boldsymbol{s_i}) \approx |\boldsymbol{r_S}| + \frac{1}{2|\boldsymbol{r_S}|}
\Bigg((\boldsymbol{r_i}\boldsymbol{S})^2-\boldsymbol{r_i}^2\Bigg) 
\end{aligned}
\end{equation}

Therefore, for ${\tau_{S}}$ in (\ref{delay}) 

\begin{equation}\label{parallax_delay3}
\begin{aligned}
\tau_{S} = &~\frac{(\boldsymbol{r_S}\boldsymbol{s_2}) - (\boldsymbol{r_S}\boldsymbol{s_1})}{c}
= \\
= &~\frac{1}{2c|\boldsymbol{r_S}|}\Bigg((\boldsymbol{r_2}\boldsymbol{S})^2-\boldsymbol{r_2}^2\Bigg) -
\frac{1}{2c|\boldsymbol{r_S}|}\Bigg((\boldsymbol{r_1}\boldsymbol{S})^2-\boldsymbol{r_1}^2\Bigg) 
\end{aligned}
\end{equation}

and, denoting $|\boldsymbol{r_S}|=r_S$

\begin{equation}\label{parallax_delay4}
\begin{aligned}
\tau_{S}= \frac{(\boldsymbol{r_2}\boldsymbol{S})^2-(\boldsymbol{r_1}\boldsymbol{S})^2-\boldsymbol{r_2}^2+\boldsymbol{r_1}^2}{2cr_S} 
\end{aligned}
\end{equation}

Equation (\ref{parallax_delay4}) provides a precise model for the parallactic delay; however, the parallax ($\pi$) is not explicitly presented here. To obtain the classical form, the baseline vector must be introduced using the equation $\boldsymbol{b} = \boldsymbol{r_{2}} - \boldsymbol{r_{1}}$.

\begin{equation}\label{parallax_delay5}
\begin{aligned}
\tau_{S} &= \frac{(\boldsymbol{r_2}\boldsymbol{S})^2-((\boldsymbol{r_2}-\boldsymbol{b})\boldsymbol{S})^2-\boldsymbol{r_2}^2+(\boldsymbol{r_2}-\boldsymbol{b})^2}{2cr_S} = \\
&= \frac{2(\boldsymbol{r_2}\boldsymbol{S})(\boldsymbol{b}\boldsymbol{S})-(\boldsymbol{b}\boldsymbol{S})^2-2(\boldsymbol{b}\boldsymbol{r_2})+\boldsymbol{|b}|^2}{2cr_S}
\end{aligned}
\end{equation}

\begin{figure}
\begin{center}

\includegraphics[scale=0.9]{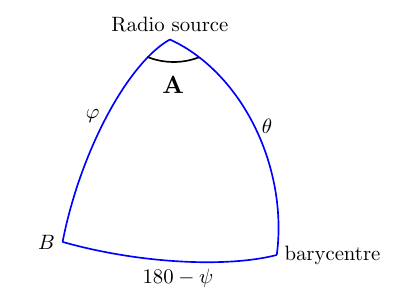} 
\vspace{0.1 cm}

\caption{Spherical triangle formed by three vectors, $\boldsymbol{s_2}$, $\boldsymbol{b}$, and $\boldsymbol{r_2}$, with respect to the Earth's geocenter. Point B lies on the celestial sphere along the extension of vector $\boldsymbol{b}$.}
\label{Fig2}
\end{center}
\end{figure}

Now, consider the spherical triangle (Fig. (\ref{Fig2})) where $\theta$ is the angle between the vectors $\boldsymbol{s_2}$ and $\boldsymbol{-r_2}$ (in this context, $\theta$ is the elongation angle between the directions from station 2 to the barycenter and the radio source),  $\varphi$ is the angle between the vectors $\boldsymbol{s_2}$ and $\boldsymbol{b}$, $\psi$ is the angle between $\boldsymbol{b}$ and $\boldsymbol{r_2}$, and angle $A$ links all the angles together through the spherical trigonometry equation:

\begin{equation}\label{parallax_delay7}
\begin{aligned}
\cos\psi = -\cos\varphi \cos\theta - \sin\theta \sin\varphi \cos A 
\end{aligned}
\end{equation}

As an extragalactic radio source is on cosmological distance, we consider that $r_{2} \ll r_{S}$ and $(\boldsymbol{r_{2}}\boldsymbol{S}) \approx (\boldsymbol{r_{2}s_{2}}) \approx (\boldsymbol{r_{2}s_{1}}) = - r_{2}\cos\theta$, therefore,  from Eq. (\ref{parallax_delay5}):
\begin{equation}\label{parallax_delay6}
\begin{aligned}
\tau_{S} = \frac{-2br_{2} \cos\varphi \cos\theta -2br_{2} \cos\psi +b^2(1-\cos^2\varphi)}{2cr_S} 
\end{aligned}
\end{equation}

and applying (\ref{parallax_delay7}):

\begin{equation}\label{parallax_delay8}
\begin{aligned}
\tau_{S} =  \frac{r_{2} \sin\theta}{r_S} \frac{b}{c}\sin\varphi \cos A + \frac{b}{c} \frac{b\sin^2\varphi}{2r_S}
\end{aligned}
\end{equation}

Now the astrometric parallax appears in the classical form $\pi = \frac{r_2}{r_S}$ in the first term of (\ref{parallax_delay8}). Denoting $\pi' = \frac{b}{2r_S}$ as an "additional" baseline parallactic displacement, if the baseline is long enough to detect the corresponding effect, one could present the final equation as follows:
 
\begin{equation}\label{parallax_delay9}
\begin{aligned}
\tau_{S} = \pi\frac{b}{c}\sin\theta \sin\varphi \cos A + \pi'\frac{b}{c} \sin^2\varphi
\end{aligned}
\end{equation}

Keeping only the first term in (\ref{parallax_delay9}), the equation for $\Delta\theta$ could be formulated in terms of the vector form

\begin{equation}\label{parallax_delay_simplified}
\begin{aligned}
\tau_{S} 
= &~\frac{(\boldsymbol{r_2}\boldsymbol{s_2})(\boldsymbol{b}\boldsymbol{s_2})-(\boldsymbol{b}\boldsymbol{r_2})}{cr_S} = \\
=& ~\frac{|r_2|}{r_S}\frac{\boldsymbol{N}(\boldsymbol{s_2}(\boldsymbol{b}\boldsymbol{s_2})-\boldsymbol{b})}{c} = \\
= &~\pi\frac{\boldsymbol{N}(\boldsymbol{s_2}(\boldsymbol{b}\boldsymbol{s_2})-\boldsymbol{b})}{c}
\end{aligned}
\end{equation}

Where $\boldsymbol{N} = \frac{\boldsymbol{r_2}}{|\boldsymbol{r_2}|} $ 
is the unit vector directed from the barycenter towards station 2. It should be noted that equation (\ref{parallax_delay9}) corresponds to the equation from the paper by \cite{Soffel_2017}, but is easier for coding.

In classical astrometric observations of parallax, an Earth-based observer measures the relative displacement of a foreground object with respect to background reference objects. For a positive meaning of the parallax, the apparent elongation angle, $\theta' = \theta + \Delta\theta$, is smaller than the elongation angle corresponding to the unperturbed position. This results in the observed object appearing shifted toward the Sun ($\Delta\theta \leq 0$) in a case of positive parallax, i.e., $\pi = -\Delta\theta$.

A hypothetical parallactic displacement in the opposite direction, where the object appears to move away from the Sun, would result in an increase in the apparent elongation angle ($\Delta\theta \geq 0$). In the classical interpretation, such a shift would imply that the foreground object is, paradoxically, further from the Sun than the background reference sources. This corresponds to a positive sign of the elongation angle change, and, inevitably, to a negative value of the annual parallax.

In classical astrometry, this scenario is unrealistic, as it suggests an inversion of the standard spatial relationships. Thus, any parallax of this nature should be referred to as "negative parallax", irrespective of its physical origin. Possible reasons for the effect's origin is discussed in two next sections.

\section{Violation of orthogonality of fundamental axes}

The orientation of the fundamental axes is defined by the positions of so-called "defining" radio sources. However, all of these radio sources are known to be active galactic nuclei (AGNs), which often exhibit extended and variable structures. Even when the structure is stable, the phase response of an interferometer depends on the baseline length and the complexity of the structure. In the worst-case scenario, the position of the same extended radio source, when measured by short and long VLBI baselines, may differ by several milliarcseconds (mas) \cite{Charlot_2020}.

Additionally, systematic effects such as secular aberration drift (SAD), with an amplitude of 5–6 $\mu$as/year (as discussed by \cite{Titov+2011}, and \cite{MacMillan+2019}), caused by the Galactocentric acceleration of the Solar System, also impact the orientation of the fundamental axes. For instance, an uncertainty in the SAD amplitude of 1 $\mu$as/year could result in a displacement of the fundamental axes by 30 $\mu$as over a 30-year period of observations \cite{Fey_2015}, \cite{Charlot_2020}. Consequently, a deviation of the axes from orthogonality by as much as $10-15$ $\mu$as is not out of the question.

Now, consider the scenario where the X and Y axes are not perfectly orthogonal. The additional term to describe the propagation of light from a radio source to the Earth along the Y-axis over the time interval $dt$ can be expressed as follows:

\begin{figure}
\begin{center}
 \includegraphics[scale=0.6]{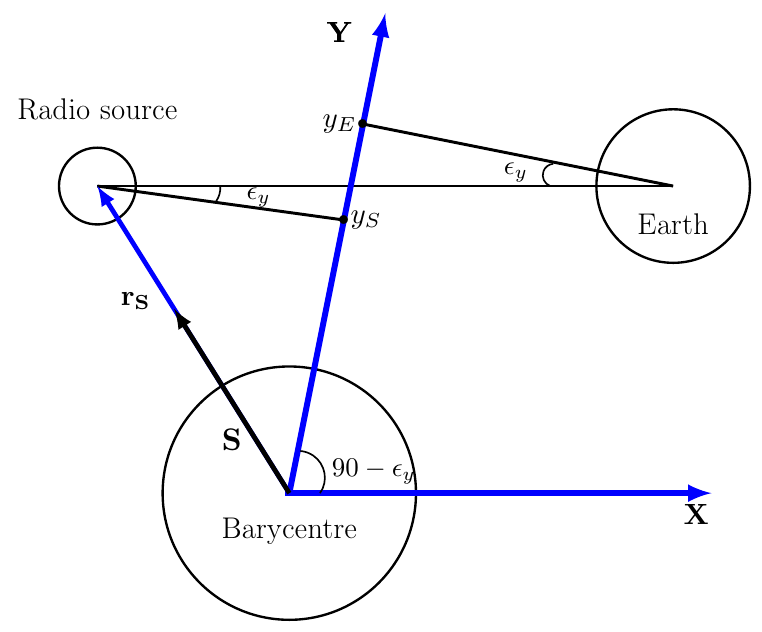} 
\vspace{0.1 cm}

\caption{Mutual position of the Earth and a radio source in the barycentric reference frame in a case if the axes X and Y are not orthogonal. The angle between X and Y is ($90- \epsilon_y$) degrees. This leads that the projecton of the signal travel path on the axes Y $y_{E} - y_{S}$ is non-zero. }

\label{Fig3}
\end{center}
\end{figure}

\begin{equation}\label{time_interval}
\begin{aligned}
cdt = dx + \epsilon_{y} dy
\end{aligned}
\end{equation}

where $\epsilon_{y}$ is a deviation of the angle between X and Y from 90 degrees, $\epsilon_{y} \ll 1$

After integration of the second term in (\ref{time_interval}):

\begin{equation}\label{time_interval2}
\begin{aligned}
c\tau_{y} =  \epsilon_{y} (y_{2} - y_{1})
\end{aligned}
\end{equation}

Since we are using a barycentric coordinate system, the projection of the signal travel path from the radio source to stations 1 and 2 onto the X and Y axes must be considered.
The projection onto the X-axis was already presented in the previous section as the difference in dot products:
$(\boldsymbol{r_S}\boldsymbol{s_2}) - (\boldsymbol{r_S}\boldsymbol{s_1}) = x_{2} - x_{1}$
as shown in equation (\ref{parallax_delay3}).

In the case where the X and Y axes are orthogonal, the difference $y_{2} - y_{1}$ should ideally be zero. However, if $\epsilon_{y} \neq 0$ (i.e., there is a small deviation), the corresponding projection along the Y-axis is given by:
$y^2_{i} =   {\boldsymbol{r_{S}}^2 - (\boldsymbol{r_{S}}\boldsymbol{s_{i}})^2}$.

From equation (\ref{dot_product2}), we have:
 $(\boldsymbol{r_S}\boldsymbol{s_i})^2 \approx r_{S}^2 + \left((\boldsymbol{r_i}\boldsymbol{S})^2  - \boldsymbol{r_i}^2 \right)$. Therefore, the time delay along the Y-axis, as described in equation (\ref{time_interval2}), can be derived accordingly.
\begin{equation}\label{t_y}
\begin{aligned}
c\tau_{y} = \epsilon_{y}(\sqrt{r^2_S -(\boldsymbol{r_S}\boldsymbol{s_2})^2} - \sqrt{r^2_S - (\boldsymbol{r_S}\boldsymbol{s_1})^2}) \approx \\
\approx \epsilon_{y} (\sqrt{\boldsymbol{r_2}^2- (\boldsymbol{r_2}\boldsymbol{S})^2} -\sqrt{\boldsymbol{r_1}^2 - (\boldsymbol{r_1}\boldsymbol{S})^2})
\end{aligned}
\end{equation}
Equation (\ref{t_y}) mirrors the form of (\ref{parallax_delay4}) from the previous section.
\begin{equation}\label{t_y2}
\begin{aligned}
\tau_{y}= \epsilon_{y}\frac{\sqrt{\boldsymbol{r_2}^2 -(\boldsymbol{r_2}\boldsymbol{S})^2} - \sqrt{\boldsymbol{r_1}^2 -(\boldsymbol{r_1}\boldsymbol{S})^2}}{c} 
\end{aligned}
\end{equation}
By substituting $\boldsymbol{r_1} = \boldsymbol{r_2} - \boldsymbol{b}$ into (\ref{t_y2}): 
\begin{equation}\label{t_y3}
\begin{aligned}
\tau_{y}= \frac{\epsilon_{y}}{c}\sqrt{\boldsymbol{r_2}^2-(\boldsymbol{r_2}\boldsymbol{S})^2} - \frac{\epsilon_{y}}{c} \sqrt{(\boldsymbol{r_2 - \boldsymbol{b}})^2 -((\boldsymbol{r_2-b})\boldsymbol{S})^2}
\end{aligned}
\end{equation}

Defining $\boldsymbol{N} = \frac{\boldsymbol{r_2}}{|\boldsymbol{r_2}|}$:

\begin{equation}\label{t_y5}
\begin{aligned}
&\tau_{y}= \frac{\epsilon_{y}|r_{2}|}{c}\sqrt{1-(\boldsymbol{N}\boldsymbol{S})^2} - 
\frac{\epsilon_{y}|r_{2}|}{c} \cdot \\
&\cdot \sqrt{1 - (\boldsymbol{N}\boldsymbol{S})^2 - 2\frac{(\boldsymbol{N}\boldsymbol{b})-(\boldsymbol{N}\boldsymbol{S})(\boldsymbol{b}\boldsymbol{S})}{|r_{2}|} + \frac{\boldsymbol{b}^2 - (\boldsymbol{b}\boldsymbol{S})^2}{|r_{2}|^2} } 
\end{aligned}
\end{equation}

After extracting $\sqrt{1 - (\boldsymbol{N}\boldsymbol{S})^2}$  from the second term and performing Taylor series expansion  (\ref{math_decomposition}) up to the second term, we obtain the following equation.

\begin{equation}\label{t_y5_1}
\begin{aligned}
 \tau_{y} = \frac{\epsilon_{y}}{c} \cdot \left( \frac{(\boldsymbol{N}\boldsymbol{b})-(\boldsymbol{N}\boldsymbol{S}) (\boldsymbol{b}\boldsymbol{S})}{
\sqrt{(1 -   (\boldsymbol{N}\boldsymbol{S})^2)}}
 -  \frac{\boldsymbol{b}^2 -(\boldsymbol{b}\boldsymbol{S})^2 }{2|r_{2}|\sqrt{(1 -  (\boldsymbol{N}\boldsymbol{S})^2)}} \right)
\end{aligned}
\end{equation}
 
As $1 - (\boldsymbol{N}\boldsymbol{S})^2 = \sin^2{\theta}$, substituting this into equation (\ref{t_y5_1}):
\begin{equation}\label{t_y6}
\begin{aligned}
\tau_{y} & = \frac{\epsilon_{y} }{c} \cdot \left(\frac{(\boldsymbol{N}\boldsymbol{b})
-(\boldsymbol{N}\boldsymbol{S})(\boldsymbol{b}\boldsymbol{S})}{\sin\theta} - \frac{\boldsymbol{b}^2-(\boldsymbol{b}\boldsymbol{S})^2 }{2|r_{2}|\sin\theta} \right)  = \\
& =  \frac{\epsilon_{y} }{c\sin\theta} \left(   (\boldsymbol{N}\boldsymbol{b})
-(\boldsymbol{N}\boldsymbol{S})(\boldsymbol{b}\boldsymbol{S}) - \frac{\boldsymbol{b}^2 -  (\boldsymbol{b}  \boldsymbol{S})^2}{2|r_{2}|} \right)  
\end{aligned}
\end{equation}

Applying the spherical triangle equations and (\ref{parallax_delay7}) we proceed:

\begin{equation}\label{delay_y1}
\begin{aligned}
\tau_{y} &= \frac{\epsilon_{y}}{c\sin\theta}\left(-b\sin\theta\sin\varphi\cos A  - \frac{\boldsymbol{b}^2 \sin^2\varphi}{2|r_2|}\right) = \\
& = -\epsilon_{y}\frac{b}{c}\sin\varphi\cos A -\epsilon_{y}\frac{\boldsymbol{b}^2 \sin^2\varphi}{2c|r_2|\sin\theta}
\end{aligned}
\end{equation}

After truncation of the small term in (\ref{delay_y1}):

\begin{equation}\label{delay_y2}
\begin{aligned}
\tau_{y} = -\epsilon_{y}\frac{b}{c}\sin\varphi\cos A 
\end{aligned}
\end{equation}

Consequently, equation (\ref{delay_y1}) bears resemblance to the parallax effect equation from the first term of (\ref{parallax_delay9}), but with two significant differences: it is independent of the distance and lacks the factor $\sin\theta$ in numerator. In other words, this apparent displacement affects all objects uniformly, regardless of their distance, and manifests as a circular motion with an amplitude of $\epsilon_{y}$, as opposed to the traditional parallactic ellipse. This effect reflects the non-orthogonality of the fundamental X and Y axes. It is evident that the observed effect can be either positive or negative, depending on the sign of the small parameter $\epsilon_y$.

\section{Implication of cosmological metrics}

Now, let us consider the stationary cosmological metric of the G{\"o}del type, as discussed in previous works (\cite{Obukhov_1992}, \cite{Korotky_1996}, \cite{Korotky_2020}):
\begin{equation}\label{Godel}
\begin{aligned}
dS^2 = c^2dt^2 - dx^2 - 2\sqrt{\sigma}e^{mx}cdtdy - ke^{2mx} dy^2  - dz^2
\end{aligned}
\end{equation}

Here, $m$, $k$, and $\sigma$ are constant parameters. In the classical G{\"o}del metric (\cite{Godel_1949}), with $m=1$, $\sigma =1$, and $k = -\frac{1}{2}$, closed timelike curves exist, leading to causality violations. However, an extended version of the metric avoids this issue if $k > 0$. The global rotation is directed along the Z-axis, with the magnitude of the rotation, $\omega$, determined by these parameters in this form:

\begin{equation}\label{rotation}
\begin{aligned}
\omega = \frac{m}{2}\sqrt{\frac{\sigma}{\sigma + k}}
\end{aligned}
\end{equation}

The vorticity vanishes when either $m$ or $\sigma$ is zero. Although further generalizations of the metric are possible (\cite{Obukhov_1992}), such cases are not essential for our current discussion.

For simplicity, let us consider the metric in the plane perpendicular to the rotation axis Z. 
\begin{equation}\label{Godel1}
\begin{aligned}
dS^2 = c^2dt^2 - 2\sqrt{\sigma}e^{mx}cdtdy - ke^{2mx} dy^2 - dx^2 - dz^2
\end{aligned}
\end{equation}

For the equation of the elementary interval $dS^2 = 0$ one could find a solution for the time interval $dt$
Solving the equation for the elementary interval $dS^2 = 0$, we can derive an expression for the time interval $dt$, assuming that $y'_x = \frac{dy}{dx} \ll 1$ and $z'_x = \frac{dz}{dx} \ll 1$.

\begin{equation}\label{Godel2}
\begin{aligned}
 &cdt   =\sqrt{\sigma}e^{mx}dy \pm \\
&~~~~\pm\sqrt{\sigma e^{2mx}dy^2 + ke^{2mx} dy^2 + dx^2 + dz^2} = \\  
&= \sqrt{\sigma}e^{mx}dy \pm \sqrt{dx^2 \bigg( 1 + (\sigma + k)  e^{2mx}\frac{dy^2}{dx^2}  + \frac{dz^2}{dx^2}\bigg)} = \\
&= \sqrt{\sigma}e^{mx}dy \pm dx\sqrt{1 + (\sigma + k) e^{2mx}y'^2_{x}  + z'^2_{x}}  \approx \\
&\approx \sqrt{\sigma}e^{mx}dy \pm dx\bigg( 1 + \frac{(\sigma + k)e^{2mx}y'^2_{x}}{2}  + \frac{z'^2_{x}}{2}\bigg) 
\end{aligned}
\end{equation}

In equation (\ref{Godel2}), we select the positive sign.
\begin{equation}\label{Godel5}
\begin{aligned}
cdt =  \bigg( 1 + \frac{(\sigma + k)e^{2mx}y'^2_{x}}{2}  + \frac{z'^2_{x}}{2}\bigg)dx + \sqrt{\sigma}e^{mx}dy 
\end{aligned}
\end{equation}

The first term in (\ref{Godel5}) corresponds to the standard geometric VLBI delay. Thus, variations in the Earth's scale factor (mean radius) could be detectable, provided that $y'_x$ and $z'_x \neq 0$. This implies that the possible global rotation of the Universe could, in principle, be detected using standard VLBI techniques, were $m$ large enough. Even if $y'_x$ and $z'_x = 0$ (i.e., all three spatial axes are orthogonal), an additional term remains, which can be expressed in the form of (\ref{time_interval}) with the parameter $\epsilon_y$.

\begin{equation}\label{Godel6}
\begin{aligned}
\epsilon_{y} = \sqrt{\sigma}e^{mx} 
\end{aligned}
\end{equation}

If we set $m = 0$ and $e^{mx} = 1$, then:

\begin{equation}\label{Godel7}
\begin{aligned}
\epsilon_{y} = \sqrt{\sigma}
\end{aligned}
\end{equation}

The additional effect is still non-zero in equation (\ref{Godel7}), even if the global rotation (\ref{rotation}) vanishes ($\omega = m =  0 $). A non-zero $\sigma$ in the metric (\ref{Godel}) is sufficient to produce a constant annual effect observable across the entire sky, independent of distance. If both parameters $\sigma$ and $m$ are non-zero, the global rotation, if it exists, would contribute to the basic effect.

\section{Discussion and conclusion}

Recent optical and VLBI observations have detected a negative parallax, drawing widespread interest due to the contentious nature of its origin. The classical parallactic delay, which includes the astrometric annual parallax, was not considered in this case for a straightforward reason: all extragalactic radio sources observed by geodetic VLBI are far too distant for an Earth-based observer to measure classical parallax, and the number of sufficiently strong galactic radio stars is limited. However, the analytical expression given in equation (\ref{parallax_delay9}) remains useful, as it explicitly illustrates the dependence on the elongation angle, $\theta$.

Interestingly, we discovered that a similar analytical dependence could arise from a more mundane cause, such as a violation of the orthogonality of the fundamental axes. In this scenario, an annual effect (though not strictly parallactic in the classical sense but rather its "circular" analog) if detected, would be independent of the object's distance. Therefore, this "quasi-parallax" effect is expected to manifest in extragalactic objects, given that the fundamental axes are determined by the coordinates of defining radio sources. Currently, there is no reliable method for independently verifying the orthogonality of these axes. Therefore, the proposed technique could be used as a handy tool to control the direct angle between the fundamental axes with sufficient accuracy. It is known that the astrometric positional stability is compromised by variations in the intrinsic structure of radio sources, with amplitudes reaching several milliarcseconds due to high-energy processes occurring in the cores of AGNs. As such, continuous monitoring of this combined instability in the fundamental axes is essential.

Furthermore, non-standard cosmological metrics, such as the classical G{\"o}del solution or a Bianchi-type III model (\ref{Godel}), might also produce an apparent annual effect. However, in these cases, the amplitude would depend on different model parameters. If the measured amplitude of the annual effect remains stable over an extended period, it could provide a basis for speculating on the estimation of specific parameters within these metrics (e.g., a global rotation of the Universe). Nonetheless, such speculation remains beyond the scope of this manuscript.

\begin{acknowledgements}

We would like to express our thanks to the Editor-in-Chief of PASA and the anonymous referee for their constructive comments and suggestions, which have significantly improved the clarity of this paper.

This paper is published with the permission of the CEO, Geoscience Australia.

\end{acknowledgements}

\bibliographystyle{pasa-mnras}
\bibliography{titov}

\clearpage
                                      
\end{document}